# Terahertz Transition-Edge Sensor with Kinetic-Inductance Amplifier at 4.2 K

Artem Kuzmin, Steffen Doerner, Stefan Singer, Ilya Charaev, Konstantin Ilin, Stefan Wuensch, and Michael Siegel

*Abstract*— **Different terrestrial terahertz applications would benefit from large-format arrays, operating in compact and inexpensive cryocoolers at liquid helium temperature with sensitivity, limited by the 300-K background radiation only. A voltage-biased Transition-Edge Sensor (TES) as a THz detector can have sufficient sensitivity and has a number of advantages important for real applications as linearity of response, high dynamic range and a simple calibration, however it requires a low-noise current readout. Usually, a current amplifier based on Superconducting Quantum-Interference Device (SQUID) is used for readout, but the scalability of this approach is limited due to complexity of the operation and fabrication. Recently, it has been shown that instead of SQUID it is possible to use a current sensor, which is based on the nonlinearity of the kinetic inductance of a current-carrying superconducting stripe. Embedding the stripe into a microwave high-Q superconducting resonator allows for reaching sufficient current sensitivity. More important, it is possible with the resonator approach to scale up to large arrays using Frequency-Division Multiplexing (FDM) in GHz range. Here, we demonstrate the operation of a voltage-biased TES with a microwave kinetic-inductance current amplifier at 4.2 K. We measured the expected intrinsic Noise-Equivalent Power NEP ~ $5\times10^{-14}$ W/Hz$^{1/2}$ and confirmed that a sufficient sensitivity of the readout can be reached in conjunction with a real TES operation. The construction of an array with the improved sensitivity ~ $10^{-15}$ W/Hz$^{1/2}$ at 4.2 K could be realized using a combination of the new current amplifier and already existing TES detectors with improved thermal isolation.**

## I. Introduction

Large-format arrays of THz detectors are required today in different terrestrial applications. Among them is THz passive security scanning for concealed hazardous objects, THz imaging for non-destructive testing (NDT) in production lines and imaging far-infrared Fourier-Transform Spectroscopy for material research and atmospheric studies [1-5].

Along with the large number of detectors in an array, these applications would benefit from the high sensitivity. The background-limited Noise-Equivalent Power (NEP) is of high importance for passive scanning applications with a stand-off THz camera [8]. High sensitivity allows to reach image acquisition speed with 25 frames per second for a required Noise Equivalent Temperature Difference (NETD ~ 0.1 K). In NDT applications a high sensitivity together with a large dynamic range will result in a higher contrast for transmission diagnostic of products. The far-infrared Fourier-Transform Spectroscopy, which employs a thermal light source, has only a limited power in the lower THz range. In this case, the use of a background-limited large-format array will allow obtaining spectral images of a sample in reasonable time. Estimations show that the sensitivity of a diffraction-limited single-pixel THz detector with Noise Equivalent Power NEP ~ $10^{-15}$ W/Hz$^{1/2}$ would be sufficient for applications with a 300-K background [6]. However, uncooled THz detectors, despite large demonstrated arrays, show only a relatively high NEP ~ $10^{-12}$ W/Hz$^{1/2}$ [7].

Cryogenic detectors can achieve the required sensitivity, but construction of large arrays is challenging. Furthermore, the cost of a cryogenic system grows rapidly for lower temperatures. A practicable system should operate at temperatures not much lower than 4.2 K, which can be reached in compact and inexpensive cryocoolers. A superconducting Transition-Edge Sensor (TES), operating in voltage-bias regime together with a low-noise current amplifier can have sufficient sensitivity at moderate temperatures [9]. Moreover, a voltage-biased TES has a strong negative electro-thermal feedback, which linearizes the response of TES and improves its dynamic range. Previously, imaging arrays with about 100 TES-bolometers were demonstrated in operation at moderately low temperatures and with the required sensitivity [10, 11]. But due to a rather complex multiplexing scheme with large number of bias and readout lines, such systems will have high costs, limited viability, and a further increase of the array size is restricted. Usually, a Superconducting Quantum Interference Device (SQUID-amplifier) is used for measurements of the current response of a TES. However, readout with many SQUID amplifiers becomes expensive and difficult in fabrication and operation. Therewith, a maximum number of wires, required for operation of SQUID-based readout, will be limited by cooling power of the cryogenic stage.

A microwave kinetic-inductance detector (MKID) is an example of a scalable detector, which can be operated in the THz spectral range. It is based on pair-breaking process in a high-Q superconducting resonator. Such approach allows for building kilo-pixel arrays with Frequency-Division

A. Kuzmin, S. Doerner, S. Singer, K. Ilin, S. Wuensch and M. Siegel are with the Institut für Mikro- und Nanoelektronische Systeme (IMS) of the Karlsruhe Institute of Technology, Karlsruhe, 76187 Germany (e-mail: artem.kuzmin@kit.edu).

I. Charaev is with the Massachusetts Institute of Technology, Cambridge, USA, was with the Institut für Mikro- und Nanoelektronische Systeme (IMS) of the Karlsruhe Institute of Technology, Karlsruhe, 76187 Germany (e-mail: charaev@mit.edu).

Multiplexing (FDM) in GHz-range using one broadband low-noise amplifier and a Software-Defined Radio (SDR) [12]. However, for a pair-breaking process to be effective in the lower THz range, the superconducting energy gap should be small. At 4.2 K this will lead to a low Q factor of the resonator and high generation-recombination noise and thus high NEP of the whole detector system. A high sensitivity at temperatures about 4.2 K was demonstrated [13] for a large array of kinetic-inductance bolometers (KIBs) [14, 15]. KIB is based on a compact lumped-element resonator, placed on a suspended THz-absorptive membrane. Due to relatively high microwave losses at these temperatures though, the resonators had required Q factor in the 100-MHz-range only. Along with relatively complex fabrication of suspended membranes, low-frequency readout limits the number of detectors per channel. Recently, we have demonstrated a THz antenna-coupled "noise" bolometer with microwave bias and readout [16]. According to experimental and theoretical estimations, it can reach the required sensitivity and a number of detectors per channel at temperature about 2 K, but it yet to be demonstrated. Additionally, it would require faster digital-signal processing at room temperatures than what was already developed for MKIDs.

In 2005, Irwin *et al.* proposed a microwave SQUID multiplexer as a readout for a voltage-biased TES [17]. It is based on a SQUID current-amplifier, embedded in a high-Q resonator as a nonlinear inductor. Many resonators with SQUIDs are then coupled to a common transmission line, which allows for FDM scheme, similar to MKID. The use of RF SQUID instead of DC SQUID in this multiplexor reduces further the number of required control and bias wires, but still, fabrication and operation of large arrays is challenging and was shown only at 100-mK temperatures for about 100 TES-detectors [18].

In 2014 Luomahaara *et al.* demonstrated a kinetic-inductance magnetometer, which utilizes the nonlinearity of the kinetic inductance of a small NbN stripe, embedded in a high-Q resonator [19]. The resonance circuit was fabricated from the same NbN layer which simplifies the fabrication, significantly. The kinetic-inductance magnetometer achieved a sensitivity comparable to SQUID and had an array scalability through FDM. Kher *et al.* presented a current amplifier based on the same principle with sensitivity ~ 5 pA/Hz$^{1/2}$ [20]. Later, it was shown, that this type of amplifier is suitable for TES which operated at 100-mK range [21].

Here, we demonstrate an operation of a voltage-biased THz TES detector together with an array-scalable current amplifier based on nonlinear-kinetic inductance at 4.2 K. The TES is an antenna-coupled superconducting nano-bolometer, and the current-sensitive inductor is a superconducting nanowire embedded in a high-Q microwave resonant circuit. The resonator is inductively coupled to a coplanar-waveguide transmission line (CPW). It allows for simultaneous injection of the TES current and measurement of the CPW transmission. We called this device a Microwave Kinetic-Inductance Nanowire Galvanometer (MKING). Previously, the MKING achieved a current sensitivity of about 10 pA/Hz$^{1/2}$ [22], which should be sufficient to reach a NEP ~ $10^{-15}$ W/Hz$^{1/2}$ with a well thermally-isolated THz TES at 4.2 K. This opens a way for building large and sensitive imaging arrays in a compact and inexpensive cryogenic system.

## II. DETECTOR SYSTEM AT 4.2 K

A detector system for demonstration of functionality consists of a single TES and a kinetic-inductance current amplifier (MKING) in separate housings. We performed measurements in a liquid-helium-bath cryostat with a THz-transparent window. The schematic of the setup is shown in Figure 1. The voltage bias of the TES is realized using a 2-Ω Manganin shunt resistor $R_{sh}$, which is biased with a current from a room-temperature source. DC lines between room and Helium temperature are filtered using cold RC filters. The current inputs of MKING are connected in series with TES. This allows ascertaining the absolute value of the current in TES. Changes of the microwave transmission of CPW in the MKING are measured using a Vector Network Analyzer (VNA). The probe signal from VNA is fed through a coaxial cable, cold attenuator and a double dc-block to the MKING block. On the other side it is connected to a cryogenic microwave low-noise high-electron-mobility transistor (HEMT) amplifier with a noise temperature $T_{n1}$ = 6 K. The second-stage amplifier is at room temperature (RT Amp. in Fig. 1) and is connected to the input of the VNA. The equivalent noise temperature $T_n$ of the complete setup is about 25 K and is dominated by the VNA, since it has only an 8-bit analog-to-digital converter (ADC).

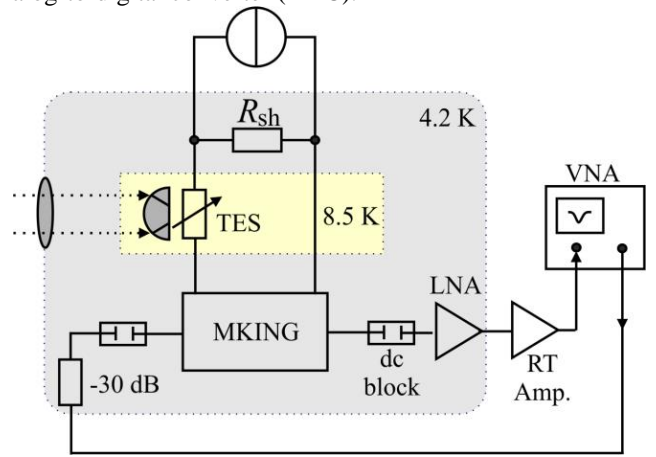

Fig. 1. The schematics of the experimental setup. The gray area is the cold stage of the 4.2 K-cryostat with THz-transparent window; the yellow area is the detector block with antenna-coupled TES on a lens at 8.5 K.

### A. Transition-Edge Sensor

We have fabricated the THz TES which is an antenna-coupled superconducting nano-bolometer from 5-nm-thin NbN film on a substrate from highly-resistive silicon with an AlN$_x$ buffer layer (Fig. 2a). The design of the TES was the same as for the NbN HEB-mixer with the log-spiral antenna for 1 – 6 THz frequency range (Fig. 1 b [23]). The antenna and contacts were patterned using electron-beam lithography and lift-off technique. The antenna is a bi-layer structure of in-situ magnetron sputtered 20-nm NbN buffer and 200-nm gold layers. To ensure good electrical contact between nano-bolometer and contacts, the developed areas of the resist mask were cleaned in-situ, using soft Ar-ion milling prior sputtering. The minimum distance of 200 nm between antenna terminals (Fig. 2c) defines the length of the nano-bolometer. The width of the nano-bolometer is 700 nm and defined using negative electron-beam resist with subsequent Ar-ion milling. The size of the nano-bolometer is adjusted so, that its normal state resistance is close to the 70-Ω impedance of the THz antenna. A preliminary characterization of the fabricated TES was done using a dc dip-stick in a transport liquid helium Dewar. Figure 3a shows the dependence of the resistance of a nano-bolometer on temperature (R-T curve). There are three superconducting transitions: the first is at $T_{c1} \approx 10.5$ K, which we attribute to the buffer layer of NbN in contacts, the second is at $T_{c2} \approx 9.4$ K, which is associated with the biggest part of the nano-bolometer and the third is around $T_{c3} = 9$ K. The transition at $T_{c3}$ is associated with the slightly degraded NbN layer of the nano-bolometer in the vicinity of contacts. For a voltage-biased TES the important parameter is $\alpha = d(\log R)/d(\log T)$ which characterizes the steepness of the transition [24]. At the point of the highest slope on the R-T curve of the TES the parameter is $\alpha \approx 40$. At temperature 4.2 K a measured I-V curve is hysteretic with a critical current $I_c = 120$ μA and with a retrapping current $I_r = 45$ μA. We also measured a non-hysteretic current-voltage characteristic (I-V curve) at 8.5 K, where electrical bias can provide a stable and uniform Joule heating of the nano-bolometer close to $T_{c2}$ (Inset in fig. 3).

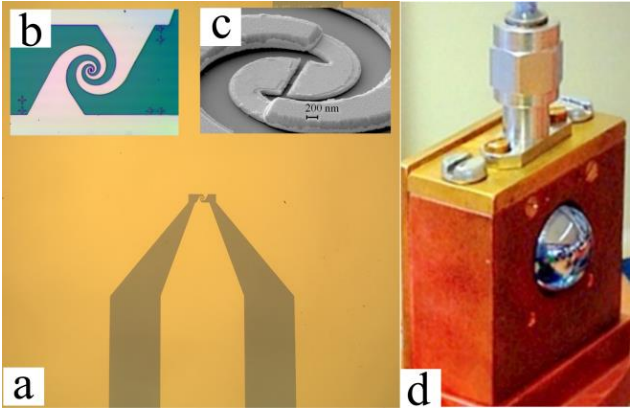

Fig. 2. The 3 mm × 3 mm chip of the antenna-coupled NbN TES (a); the log-spiral planar antenna of the TES for 1-6 THz range (b); the central part of the antenna with the NbN nano-bolometer (c); the detector block with a silicon lens and a built-in bias-T (d).

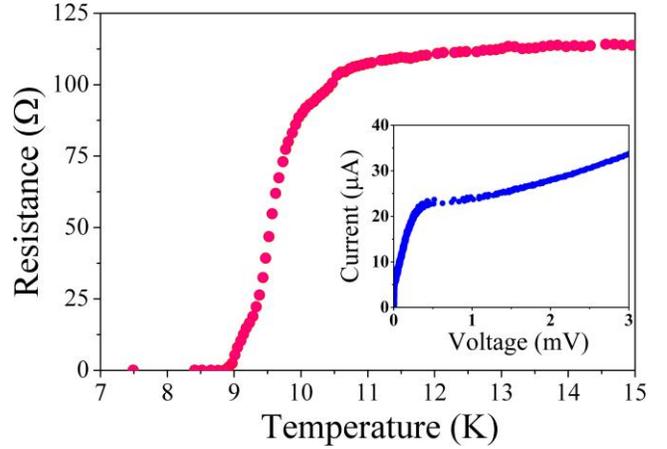

Fig. 3. The R-T curve of the nano-bolometer. Inset: the I-V curve of the nano-bolometer measured at $T = 8.5$ K in a current-bias mode.

From this I-V curve we estimated a thermal conductance of the nano-bolometer $G_{th} \approx 25$ nW/K and a loop gain for the negative electro-thermal feedback $L = \alpha I V_b / G_{th} T \approx 4$ for a working point $V_b = 1$ mV and $T = 9.4$ K. Using these values, we calculated the current responsivity $\Re_I = (L/L+1)/V_b \approx 800$ A/W, a minimum phonon-limited $NEP_{ph} = \sqrt{4kT^2 G_{th}} \approx 10^{-14}$ W/Hz$^{1/2}$, and a minimum noise current $\delta I \approx 8$ pA/Hz$^{1/2}$ of the nano-bolometer. The TES chip with the size of 3 mm × 3 mm was glued in the focus of the hyper-hemispherical silicon lens with 12-mm diameter and without anti-reflection coating. The lens with the chip were mounted in a copper block with an adapter plate (Fig. 1d), which has a built-in bias-T and allows for a complete galvanic decoupling of the TES from the cryostat. The temperature sensor and a heater were integrated into a detector block for the control of the TES operation temperature $T$.

*B. Microwave Kinetic-Inductance Nanowire Galvanometer*

To avoid in the future a limited scalability and complexity of readout based on SQUID-amplifier, we replaced it with a superconducting current amplifier, MKING. The circuit design of the MKING is shown in Figure 4 and is based on planar lumped-elements. The complete device is fabricated from a single 5-nm-thick layer of NbN on a sapphire substrate using electron-beam lithography and ion milling. The nonlinear inductor $L_k$ is a 100-nm-wide straight nanowire (Fig. 5, inset). Together with the interdigitated capacitor $C$, it forms a high-Q microwave resonator, which is on one side directly connected to a ground plane. On the other side the resonator is inductively coupled to the center of the CPW feed line using a 1-μm-wide inductor $L_c$. A measured critical current of inductor line is $I_c = 60$ μA at 4.2 K, which is at least a factor 3 higher than the current of the TES in the working point. The resistor $R_{diss}$, in series with $L_k$, represents microwave dissipation at the resonance frequency. It will be shown later, that it is the nanowire, which mainly contributes to the dissipation. The resonance frequency of the device is about 4.7 GHz. The response of the device is the change of the complex voltage transmission $S_{21}$ from port 1 to port 2 of the CPW feed line measured with a microwave generator at port 1 with output impedance $Z_0$ =50 Ω and a receiver with the same impedance at port 2 (Figure 4). In our experiment we used phase response of the device with a fixed frequency of the microwave probe. The single chip of MKING device with a size of 3 mm × 3 mm was mounted into a gold-plated brass block with an adapter plate, two SMA connectors for CPW feed line and a DC input port (Figure 5b). The chip was bonded to the adapter plate using aluminum wires. The particular MKING device had a noise current of about 40 pA/Hz$^{1/2}$ (Figure 6), which is higher than the value that was demonstrated with the previous 3.8-GHz device [22]. An excess 1/f noise in the spectrum (Figure 5) is due to the lack of magnetic shielding and insufficient filtering.

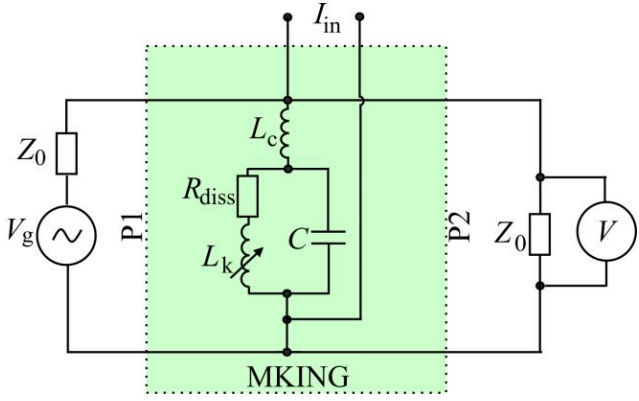

Fig. 4. The circuit design of the single MKING device. A matched microwave generator is on port 1 (P1); a matched receiver is on port 2 (P2). The $R_{diss}$ represents microwave dissipation in the resonator.

### III. MEASUREMENT OF RESPONSIVITY AND NEP

The pre-calibrated shift of MKING's resonance frequency can be used to measure IV-curves of the TES in a wide range of voltages directly in the setup with the cryostat (Fig. 1). The pre-calibration was conducted separately in a DC/RF dip-stick and VNA in a frequency-sweep resonance-track mode.

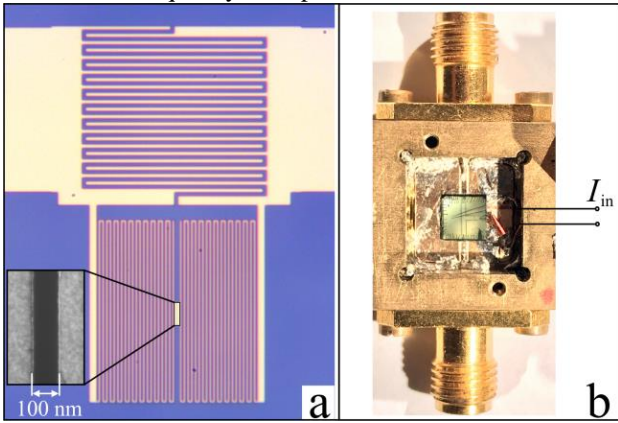

Fig. 5. (a) Optical image of the MKING resonator; inset: the scanning-electron microscope image of the nonlinear nanowire inductor; (b) the microwave block with the single MKING (in the center), SMA connectors and a dc input $I_{in}$.

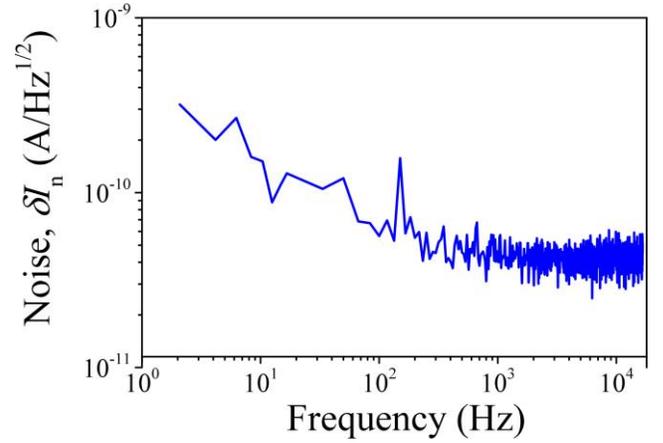

Fig. 6. Measured noise spectrum of the MKING device.

We obtained a set of I-V curves at different operation temperatures of the TES (Fig. 7). From these I-V curves we verified the electrical responsivity $\Re_I \approx 700$ A/W at $V_b$ = 1 mV and $T$ = 8.5 K, and a thermal conductivity which are close to the value obtained from the dc current-biased measurement. To measure the optical response, to check its linearity and define a dynamic range of the TES, we applied a THz signal from a calibrated 0.65-THz quasi-optical source through the cryostat window. The dependence of the current trough TES on applied THz power is shown in Figure 8. The measured optical responsivity of 3.3 A/W is much lower than the electrical one due to the low coupling efficiency $\eta = 0.5\%$, since we have a not optimized matching of Gaussian beams from the THz source and TES antenna. Moreover, antenna was not designed for frequencies below 1 THz and its performance at 0.65 THz might be low.

The dynamic range, where the response of the TES is linear, reaches >50 dB, if measured using the frequency sweep of VNA. For a real application with many detectors a frequency sweep is not possible. In case of a fixed-frequency probe, the dynamic range is limited by the dynamic range of the MKING, which is around 30 dB.

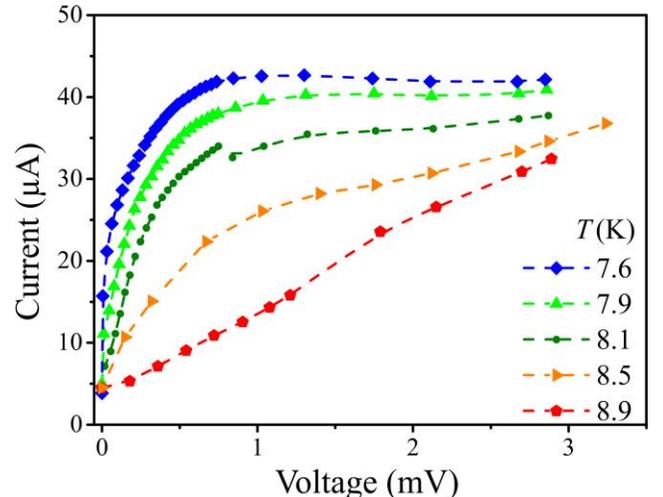

Fig. 7. Set of I-V curves of the TES measured with MKING in the cryostat at different temperatures of the block indicated in the legend.

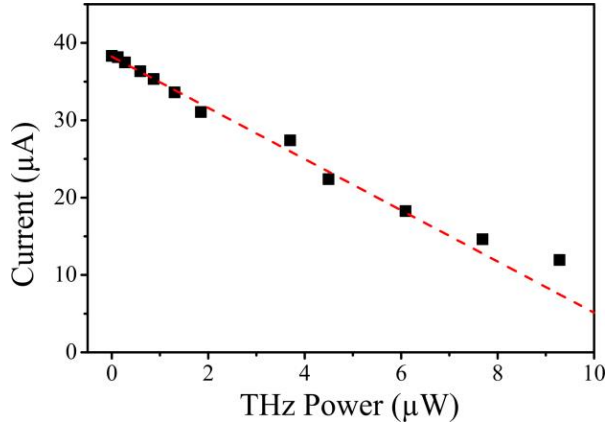

Fig. 8. Current response of the TES on THz signal. The dashed line is a linear fit. The data are obtained using frequency sweep and resonance tracking on VNA.

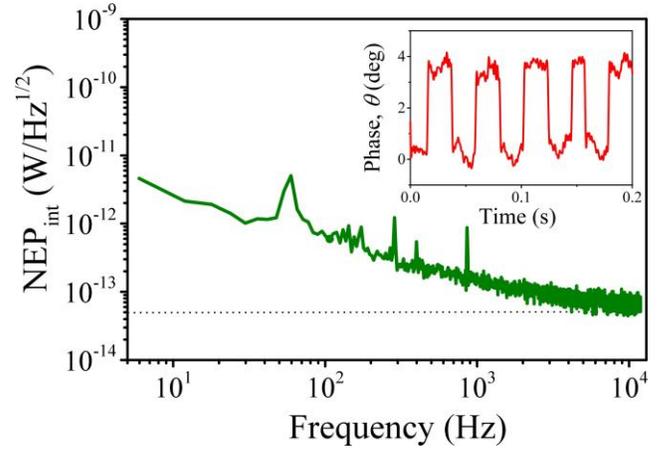

Fig. 9. Measured intrinsic NEP of the detector system. The phase response of the TES-MKING to 160 pW of absorbed THz power (in the inset).

To measure a small-signal response, we applied a chopped THz signal, which corresponds to an absorbed power of $P_{abs} = 160$ pW. A time trace of the phase response of the MKING to a changing current through TES is shown in the inset of Figure 9. The intrinsic phase-watt responsivity of $\Re_{\theta-W} \approx 2 \times 10^{10}$ deg/W was obtained in this measurement. To determine intrinsic NEP, we recorded a time trace of the phase without THz signal and plotted spectrum of the phase noise, divided by the measured $\Re_{\theta-W}$ (Fig. 9).

## IV. DISCUSSION

The measured intrinsic NEP reaches $5 \times 10^{-14}$ W/Hz$^{1/2}$ at white noise level, which is a factor 5 higher than the estimated value. It is probably due to the prevalence of the noise current of the particular MKING over noise of the TES. Indeed, if we take the separately measured TES responsivity and the noise current of the MKING we end up with the same NEP. A reduction of the effective noise temperature of the microwave readout could be achieved using back-end electronics with ADC of higher resolution. This would result in a factor 2 lower noise current of the MKING which is limited by the phase noise of the HEMT amplifier (phase noise of the amplifier is $\delta\theta_{amp} = \sqrt{k_B T_n / 2 P_{in}}$, $P_{in}$ is a RF power on the input). Additionally, using the previous version of MKING with lower critical current [22] it would be possible to reach a factor 4 lower NEP.

In order to obtain system NEP ~ $10^{-15}$ W/Hz$^{1/2}$, which is required for the 300-K background-limited operation, the performance of TES and MKING should be improved.

### A. TES optimization

A system NEP ~ $10^{-15}$ W/Hz$^{1/2}$ at 4.2 K would require the following parameters of the TES:

a) thermal conductance $G_{th}$ ~ 1 nW/K at the operation temperature $T \approx T_{bath} = 4.2$ K for sufficiently low phonon noise NEP$_{ph}$;

b) current responsivity $\Re_I \approx 1/V_b \geq 10^4$ A/W is required to overcome noise of readout (typically $\delta I_n \approx 10$ pA/Hz$^{1/2}$).

If we define a current at working point to be around $I_b = 15$ μA (about $0.5 I_c$ of a typical MKING) and a THz load from 300-K background in the range 350±50 GHz to be $P_b \approx 200$ pW [25], then:

c) a critical temperature of a TES $T_c \approx T + (I_b V_b + P_b)/G_{th} \approx 5.9$ K and a dc resistance of the TES at working point $R_{wp} \approx V_b / I_b \approx 7$ Ω;

d) for the strong ETF the loop gain $L$ should be $L \approx I_b^2 dR / G_{th} dT \geq 10$; it requires a derivative $dR/dT \geq 45$ Ω/K at the working point on the superconducting transition.

For bolometers on a SiN$_x$ membrane the necessary $G_{th}$ has been already demonstrated [26, 13] at moderately low temperatures (1 – 5 K). A TES with suitable $T_c$ and a slope on R-T curve could be fabricated, for example, from a 10-nm-thin niobium film with reduced critical temperature due to the intrinsic proximity effect [27].

Alternatively, the required parameters can be achieved for antenna-coupled suspended superconducting nano-bolometers which already demonstrated sufficient sensitivity [28]. In this case, large membranes are not required and fabrication process is less complex. Both, quasi-optical-lens and waveguide coupling could be used for this type of TES.

### B. MKING optimization

Beside the necessary noise current, there are following prerequisites for MKING:

a) resonance frequency $f_r$ should be in the range of 4-8 GHz, where broadband low-noise HEMT amplifiers are available;

b) bandwidth of a single device $\Delta f \approx$ BW$/N_{det} \approx 5$ MHz, where $N_{det} \geq 100$ is the number of detectors per readout line, BW is the available modulation bandwidth for FDM (typically 1-2 GHz);

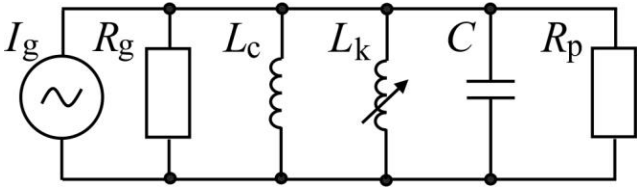

Fig. 10. Norton equivalent circuit of the MKING in the form of parallel LCR-resonator.

c) loaded Q factor $Q_L = f_r / \Delta f \approx 500$ follows from a) and b); coupling and internal Q factors: $Q_c, Q_i > Q_L$, since they are connected through the relation $Q_L = (1/Q_c + 1/Q_i)^{-1}$;

d) minimum width $w$ and thickness $d$ of inductors are limited by the fabrication technology and by values of necessary critical current (typically for NbN $w = 100$ nm, $d = 5$ nm, $I_c \approx 30$ µA);

e) values of the inductances $L_c, L_k$ and capacitance $C$ are dependent on each other due to required $f_r$;

f) maximal values of MKING inductors $L_c, L_k$ and $C$ should comply with lumped-element requirements to avoid self-resonance [29].

g) Additionally, a stability criterion for TES voltage-bias with strong ETF requires that electro-thermal time constant should be significantly shorter than thermal time constant of the TES [30]. The latter limits value of total inductance to $L_c + L_k \leq R_{wp} C_{th} / 6 G_{th}$, where $C_{th}$ is the thermal capacitance of a TES.

To understand how parameters of the MKING influence noise, we analyze its small-signal responsivity by considering Norton equivalent of the MKING circuit (Fig. 10).

In this scheme $R_g(\omega) = 2(\omega L_c)^2 / Z_0$ is the generator output impedance and $R_p(\omega) = (\omega L_k)^2 / R_{diss}$ is a parallel equivalent resistance.

The small-signal phase responsivity is the product of three derivatives:

$$\Re_\theta = \frac{d\theta}{d\omega} \times \frac{d\omega}{dL_k} \times \frac{dL_k}{dI} \quad (1)$$

Near resonance, the phase angle is $\theta \approx \text{Im}(S_{21}) / S_{21,m}$, where $S_{21,m}$ is a minimum voltage transmission at resonance. Its frequency dependence is given by:

$$S_{21}(\omega) = \frac{Q_L / Q_i + 2jQ_L x}{1 + 2jQ_L x} \quad (2)$$

where $Q_i = \omega_r R_p C$ and $Q_L = \omega_r R_{tot} C$ are intrinsic and loaded quality factors respectively, $\omega_r = 1/\sqrt{L_{tot} C}$ is the resonance frequency, $x = (\omega - \omega_r)/\omega_r$ is the fractional frequency shift near resonance, $R_{tot} = R_g \| R_p$ and $L_{tot} = L_c \| L_k$ are total resistance and inductance of parallel elements respectively. The kinetic-inductance nonlinearity can be derived from Ginzburg-Landau or Bardeen-Cooper-Schrieffer theories, for example like in reference [31]. The nonlinearity can be described by the universal normalized derivative $\varepsilon(i) = (dL_k / dI) \times (I_c / L_k)$, where $i = I / I_c$ is the relative current. For the range of currents, which corresponds to $i = 0.2 \div 0.7$ this derivative could be approximated as $\varepsilon(i) \approx 0.22 i$. Combining $\varepsilon(i)$ with (1) and (2), we derive the phase responsivity:

$$\Re_\theta = \frac{\varepsilon(i)}{I_c} \cdot \frac{Q_L^2}{Q_c} \cdot \frac{1}{(1+\gamma) S_{21,m}} \quad , \quad (3)$$

where $\gamma = L_k / L_c$. The noise current of the MKING device is a quadrature sum of the internal noise current and noise from LNA. The phase noise of the LNA is $\delta\theta_{amp} = \sqrt{kT_n / 2 S_{21,m}^2 P_g}$, where $T_n$ is the noise temperature, $P_g = V_g^2 / 4 Z_0$ is the generator output power. Dividing phase noise by responsivity from (3), we obtain an equivalent noise current of the MKING due to the LNA:

$$\delta I_{amp} = \frac{I_c}{\varepsilon(i)} \frac{Q_c}{Q_L^2} \sqrt{\frac{kT_n}{2 P_g}} (1+\gamma) . \quad (4)$$

From (4) it follows that, an increase of the microwave power $P_g$ would result in a lower value of $\delta I_{amp}$, but there are limitations. The relative current $i$ in a nanowire, in this case, is the sum of the dc and microwave currents $i = i_{dc} + i_{RF}$. In figure 11 one can see dependences of the loaded Q factor $Q_L$ and $R_{diss}$ on current for our particular MKING. The $Q_L$ degrades rapidly due to increasing losses with increasing current and thus decreasing internal Q factor $Q_i$. Due to constraints set by necessary $Q_L$, losses will limit maximum relative current $i$, and thus, a maximum probe power $P_g$. But besides that, there is a resonance bifurcation phenomenon [32], which limits generator power to a maximum value $P_{g,m} \approx \omega L_k (I_c / Q_L)^2 \times (Q_c / Q_L)$, or a maximum fraction of the RF current to the value $i_{RF,m} \approx 1/\sqrt{Q_L (1+\gamma)} \leq 5\%$. If we plug $P_{g,m}$ into (4), then noise current is:

$$\delta I_{amp} \approx 4.5 \frac{I_c}{I_{in}} \sqrt{\frac{1}{Q_L}\left(\frac{kT_n}{Z_0}\right)} \frac{(1+\gamma)^{3/2}}{\gamma} \quad (5)$$

Here, the relative current $i = I_{in} / I_c$ should be matched to TES bias and maximized until $Q_i(i)$ drops to a value, which corresponds to $Q_L = 500$.

From (5) it is seen, that $\gamma = L_k / L_c$ should be as high as possible to get low $\delta I_{amp}$. The only free parameter, which is left to define parameter $\gamma$, is the value of the coupling inductor $L_c$. The increase of $L_c$ allows larger $\gamma$ and lead to improvement of the noise current, but only to a certain extent. Larger values of inductances, in this case, could only be achieved by the increase of inductor length (nanowire cross section is fixed by the necessary $I_{in}$). The latter will lead to a bigger inductor volume and thus, higher generation-recombination noise of the MKING which will dominate over noise of the amplifier. The estimation with realistic parameters for our NbN nanowires shows that, the noise current of MKING could be improved to values of $\delta I_{min} \approx 2\text{-}3$ pA/Hz$^{1/2}$, which should be sufficient for TES readout.

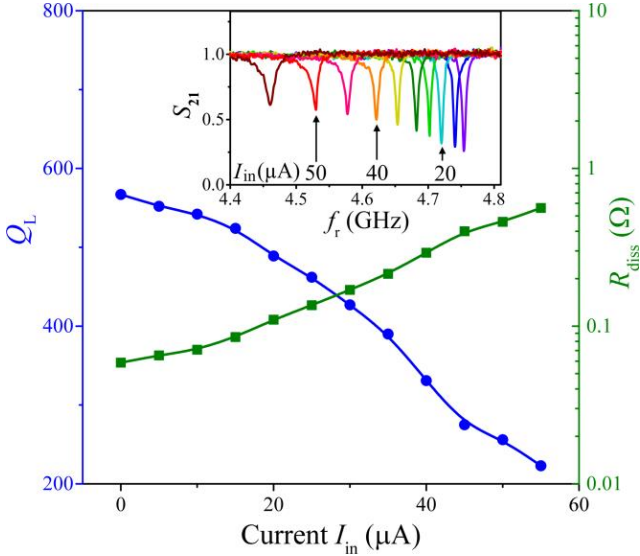

Fig. 11. Dependences of loaded Q-factor $Q_L$ and the equivalent resistance $R_{diss}$ in the MKING on the input current. Inset: raw data - set of resonance curves at different input currents indicated in the graph.

The reduction of the 1/f noise of the MKING could be done using additional magnetic shielding of the detector block with mu-metal along with a differential or "gradiometric" design of the single current amplifier (Fig. 12). In this case two MKINGs are used for measuring current in TES. If the responsivities of both galvanometers are nearly equal, then by measuring a sum of phase shifts one can effectively reject common mode interference

*C. Multiplexing scheme*

A suggested multiplexing scheme for an array of voltage-biased TES detectors is shown in figure 12. All TES detectors are biased in parallel using one current source and individual shunt resistors $R_{sh}$. MKING devices are separated by the small inline resistors $R_{ser}$, which prevents leakage of the TES current into neighbor current amplifier. In total, the array requires one dc line, one microwave line and one broadband HEMT LNA for operation. Frequency-division multiplexing of the array can be realized in the same way as for KIDs with existing RF boards and digital signal processing [33]. The achieved loaded Q factor of the MKING $Q_L \sim 500$ allows placing ~100 detectors in the 1-GHz modulation bandwidth.

## V. CONCLUSION

The detector system for demonstration of functionality with voltage-biased antenna-coupled TES and kinetic-inductance current amplifier at temperature 4.2 K reached internal NEP $\sim 5 \times 10^{-14}$ W/Hz$^{1/2}$, which is close to the estimated phonon-limited NEP of the particular TES. The sensitivity of the system is limited by the particular current amplifier and the microwave setup. It could be further improved by a factor 2 using baseband ADC with higher resolution. Our analysis shows that NEP $\sim 10^{-15}$ W/Hz$^{1/2}$ is feasible with already existing well-isolated THz TES and MKING with improved design and back-end electronics. The obtained values of Q factor of resonator in current amplifier allow to scale up detector system to >100 pixel per readout channel at 4.2 K.

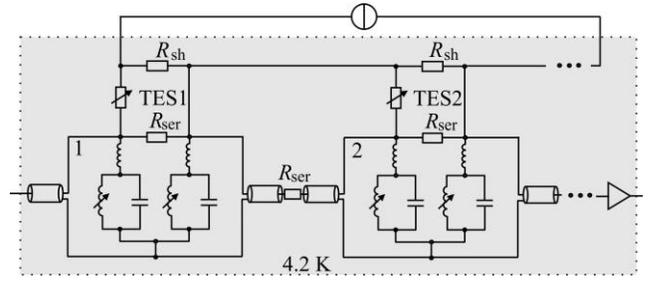

Fig. 12. Schematics of an array of voltage-biased TESs with a differential MKINGs; the dark area is a cryogenic stage at $T = 4.2$ K.

ACKNOWLEDGMENT

Authors would like to thank M. Schmelz and R. Stolz (IPHT Jena) for the fruitful discussions and help during measurements, A. Stassen and K.-H. Gutbrod for the help with preparation of samples and for fabrication of mechanical parts.

REFERENCES

[1] J. P. Guillet, B. Recur, L. Frederique, B. Bousquet, L. Canioni, I. Manek-Hönninger, P. Desbarats, and P. Mounaix, "Review of Terahertz Tomography Techniques," *Journal of Infrared, Millimeter, and Terahertz Waves*, vol. 35, no. 4, pp. 382–411, 2014.
[2] E. Bründermann, H.-W. Hübers, M. F. Kimmitt, Terahertz Techniques New York, Springer, p 310, 2012.
[3] T. Robin, C. Bouye, and J. Cochard, "Terahertz applications: trends and challenges," *Terahertz, RF, Millimeter, and Submillimeter-Wave Technology and Applications VII*, Jul. 2014.
[4] P. Dean, A. Valavanis, J. Keeley, K. Bertling, Y. L. Lim, R. Alhathlool, A. D. Burnett, L. H. Li, S. P. Khanna, D. Indjin, T. Taimre, A. D. Rakić, E. H. Linfield, A. G. Davies, "Terahertz imaging using quantum cascade lasers—a review of systems and applications", *J. Phys. D: Appl. Phys*. 47, p. 374008, 2014.
[5] H. Latvakoski, M.G. Mlynczak, D.G. Johnson, R.P. Cageao, D.P. Kratz, K. Johnson, "Far-infrared spectroscopy of the troposphere: instrument description and calibration performance", *Applied optics*, 52(2), pp.264-273, 2013.
[6] E. Heinz, T. May, D. Born, G. Zieger, S. Anders, G. Thorwirth, V. Zakosarenko, M. Schubert, T. Krause, M. Starkloff, A. Krüger, M. Schulz, F. Bauer, and H.-G. Meyer, "Passive Submillimeter-wave Stand-off Video Camera for Security Applications," *Journal of Infrared, Millimeter, and Terahertz Waves*, vol. 31, no. 11, pp. 1355–1369, 2010.
[7] S. V. Berkel, O. Yurduseven, A. Freni, A. Neto, and N. Llombart, "THz Imaging Using Uncooled Wideband Direct Detection Focal Plane Arrays," *IEEE Transactions on Terahertz Science and Technology*, vol. 7, no. 5, pp. 481–492, 2017.
[8] E. Heinz, T. May, D. Born, G. Zieger, A. Brömel, S. Anders, V. Zakosarenko, T. Krause, A. Krüger, M. Schulz, F. Bauer, and H.-G. Meyer, "Development of passive submillimeter-wave video imaging systems for security applications," *Millimetre Wave and Terahertz Sensors and Technology V*, 2012.
[9] A. Luukanen and J. P. Pekola, "A superconducting antenna-coupled hot-spot microbolometer," *Applied Physics Letters*, vol. 82, no. 22, pp. 3970–3972, Feb. 2003.
[10] E. Heinz, T. May, D. Born, G. Zieger, K. Peiselt, V. Zakosarenko, T. Krause, A. Krüger, M. Schulz, F. Bauer, and H.-G. Meyer, "Progress in passive submillimeter-wave video imaging," *Passive and Active Millimeter-Wave Imaging XVII*, Sep. 2014.
[11] A. Luukanen, T. Kiuru, M. M. Leivo, A. Rautiainen, and J. Varis, "Passive three-colour submillimetre-wave video camera," *Passive and Active Millimeter-Wave Imaging XVI*, 2013.
[12] P. K. Day, H. G. Leduc, B. A. Mazin, A. Vayonakis, and J. Zmuidzinas, "A broadband superconducting detector suitable for use in large arrays," *Nature*, vol. 425, no. 6960, pp. 817–821, 2003.
[13] A. Timofeev, J. Luomahaara, L. Gronberg, A. Mayra, H. Sipola, M. Aikio, M. Metso, V. Vesterinen, K. Tappura, J. Ala-Laurinaho, A. Luukanen, and J. Hassel, "Optical and Electrical Characterization of a Large Kinetic Inductance Bolometer Focal Plane Array," *IEEE*

[14] A. V. Timofeev, V. Vesterinen, P. Helistö, L. Grönberg, J. Hassel, and A. Luukanen, "Submillimeter-wave kinetic inductance bolometers on free-standing nanomembranes," *Superconductor Science and Technology*, vol. 27, no. 2, p. 025002, Dec. 2013.

[15] M. Arndt, S. Wuensch, A. Kuzmin, S. Anders, T. May, M. Schubert, H.-G. Meyer, and M. Siegel, "Novel Detection Scheme for Cryogenic Bolometers with High Sensitivity and Scalability," *IEEE Transactions on Applied Superconductivity*, vol. 25, no. 3, pp. 1–5, 2015.

[17] K. D. Irwin and K. W. Lehnert, "Microwave SQUID multiplexer," *Applied Physics Letters*, vol. 85, no. 11, pp. 2107–2109, 2004.

[18] B. Dober, D. T. Becker, D. A. Bennett, S. A. Bryan, S. M. Duff, J. D. Gard, J. P. Hays-Wehle, G. C. Hilton, J. Hubmayr, J. A. B. Mates, C. D. Reintsema, L. R. Vale, and J. N. Ullom, "Microwave SQUID multiplexer demonstration for cosmic microwave background imagers," *Applied Physics Letters*, vol. 111, no. 24, p. 243510, Nov. 2017.

[19] J. Luomahaara, V. Vesterinen, L. Grönberg, and J. Hassel, "Kinetic inductance magnetometer," *Nature Communications*, vol. 5, no. 1, Oct. 2014.

[20] A. Kher, P. K. Day, B. H. Eom, J. Zmuidzinas, and H. G. Leduc, "Kinetic Inductance Parametric Up-Converter," *Journal of Low Temperature Physics*, vol. 184, no. 1-2, pp. 480–485, 2015.

[21] A. S. Kher, "Superconducting Nonlinear Kinetic Inductance Devices", Ph. D. thesis, California Institute of Technology, Pasadena, USA, 2017.

[22] S. Doerner, A. Kuzmin, K. Graf, I. Charaev, S. Wuensch, and M. Siegel, "Compact microwave kinetic inductance nanowire galvanometer for cryogenic detectors at 4.2 K," *Journal of Physics Communications*, vol. 2, no. 2, p. 025016, 2018.

[23] H. Richter, S. G. Pavlov, A. Semenov, M. Wienold, L. Schrottke, K. Biermann, R. Hey, H.-T. Grahn, K. Ilin, M. Sigel, and H.-W. Hübers, "Development of a 4.7-THz front end for the GREAT heterodyne spectrometer on SOFIA," 23rd International Symposium on Space Terahertz Technology, 2.-4. Apr. 2012, Tokyo, *eLib*, 02-Apr-2012. [Online]. Available: https://elib.dlr.de/78698/.

[24] K. D. Irwin, "An application of electrothermal feedback for high resolution cryogenic particle detection," *Applied Physics Letters*, vol. 66, no. 15, pp. 1998–2000, Oct. 1995.

[25] T. May, V. Zakosarenko, E. Kreysa, W. Esch, S. Anders, H.-P. Gemuend, E. Heinz, and H.-G. Meyer, "Design, realization, and characteristics of a transition edge bolometer for sub-millimeter wave astronomy," *Review of Scientific Instruments*, vol. 83, no. 11, p. 114502, 2012.

[26] M. Arndt, S. Wuensch, C. Groetsch, M. Merker, G. Zieger, K. Peiselt, S. Anders, H.G. Meyer, M. Siegel, "Optimization of the Microwave Properties of the Kinetic-Inductance Bolometer (KIBO)" *IEEE Trans. Appl. Supercond.*, 27(4), pp.1-5, 2017.

[27] K. Ilin, S. Vitusevich, B. Jin, A. Gubin, N. Klein, and M. Siegel, "Peculiarities of the thickness dependence of the superconducting properties of thin Nb films," *Physica C: Superconductivity*, vol. 408-410, pp. 700–702, 2004.

[28] P. Helisto, J. Penttila, H. Sipola, L. Gronberg, F. Maibaum, A. Luukanen, and H. Seppa, "NbN Vacuum Bridge Bolometer Arrays With Room Temperature Readout Approaching Photon Noise Limited THz Imaging Applications," *IEEE Transactions on Applied Superconductivity*, vol. 17, no. 2, pp. 310–313, 2007.

[29] D. F. Santavicca, J. K. Adams, L. E. Grant, A. N. Mccaughan, and K. K. Berggren, "Microwave dynamics of high aspect ratio superconducting nanowires studied using self-resonance," *Journal of Applied Physics*, vol. 119, no. 23, p. 234302, 2016.

[30] K. D. Irwin, G. C. Hilton, D. A. Wollman, and J. M. Martinis, "Thermal-response time of superconducting transition-edge microcalorimeters," *Journal of Applied Physics*, vol. 83, no. 8, pp. 3978–3985, 1998.

[31] A. J. Annunziata, D. F. Santavicca, L. Frunzio, G. Catelani, M. J. Rooks, A. Frydman and D. E. Prober, "Tunable superconducting nanoinductors" *Nanotechnology*, 21, 44, p. 445202, 2010.

[32] L. J. Swenson, P. K. Day, B. H. Eom, H. G. Leduc, N. Llombart, C. M. Mckenney, O. Noroozian, and J. Zmuidzinas, "Operation of a titanium nitride superconducting microresonator detector in the nonlinear regime," *Journal of Applied Physics*, vol. 113, no. 10, p. 104501, 2013.

[33] J. V. Rantwijk, M. Grim, D. V. Loon, S. Yates, A. Baryshev, and J. Baselmans, "Multiplexed Readout for 1000-Pixel Arrays of Microwave Kinetic Inductance Detectors," *IEEE Transactions on Microwave Theory and Techniques*, vol. 64, no. 6, pp. 1876–1883, 2016.